\documentclass[conference, a4paper]{IEEEtran}
\IEEEoverridecommandlockouts

\usepackage[font=footnotesize,caption=false]{subfig} 

\usepackage{cite}
\usepackage{amsmath,amssymb,amsfonts, amsthm}
\usepackage{algorithmic}
\usepackage{graphicx}
\usepackage{textcomp}
\usepackage{xcolor}
\usepackage{booktabs}
\usepackage{glossaries}
\usepackage{subfig}
\usepackage{algorithm}
\usepackage{algorithmic}
\usepackage{bbm}
\usepackage{bm}
\usepackage{multirow}
\usepackage{tabularx}
\usepackage{amsmath}

\usepackage{tikz}
\usetikzlibrary{shapes,arrows}
\usetikzlibrary{plotmarks}
\usetikzlibrary{patterns}
\usetikzlibrary{arrows.meta}
\usetikzlibrary{positioning}
\usetikzlibrary{spy}

\usepackage{pgfplots}
\usepgfplotslibrary{colorbrewer}
\usepgfplotslibrary{fillbetween}

\def\BibTeX{{\rm B\kern-.05em{\sc i\kern-.025em b}\kern-.08em
    T\kern-.1667em\lower.7ex\hbox{E}\kern-.125emX}}

\newacronym{aoi}{AoI}{Age of Information}
\newacronym{irsa}{IRSA}{irregular repetition slotted ALOHA}
\newacronym{sic}{SIC}{successive interference cancellation}
\newacronym{iot}{IoT}{Internet of Things}
\newacronym{nbiot}{NB-IoT}{Narrowband IoT}
\newacronym{sinr}{SINR}{signal-to-interference-plus-noise ratio}
\newacronym{plr}{PLR}{packet loss rate}
\newacronym{ne}{NE}{Nash equilibrium}
\newacronym{esu}{ESU}{expected number of successful users}
\newacronym{bs}{BS}{base station}
\begin{document}
\title{A Game-Theoretic Perspective for Efficient Modern Random Access}
\author{\IEEEauthorblockN{Andreas Peter Juhl Hansen\IEEEauthorrefmark{1}, Jeppe Roden Münster\IEEEauthorrefmark{1}, Rasmus Erik Villadsen\IEEEauthorrefmark{1}, Simon Bock Segaard\IEEEauthorrefmark{1},\\ Søren Pilegaard Rasmussen\IEEEauthorrefmark{1}, Christophe Biscio\IEEEauthorrefmark{1}, and Israel Leyva-Mayorga\IEEEauthorrefmark{2}}\IEEEauthorblockA{\IEEEauthorrefmark{1}Department of Mathematical Sciences, Aalborg University, Denmark\\ (\{apjh21, jmunst21, revi21, ssegaa21, sorras21\}@student.aau.dk, christophe@math.aau.dk)\\\IEEEauthorrefmark{2}Department of Electronic Systems, Aalborg University, Denmark (ilm@es.aau.dk).}\thanks{A. P. J. Hansen, J. R. Münster, R. E. Villadsen, S. B. Segaard, and S. P. Rasmussen are co-first authors.}}
\maketitle
\begin{abstract}
Modern random access mechanisms combine packet repetitions with multi-user detection mechanisms at the receiver to maximize the throughput and reliability in massive \gls{iot} scenarios. However, optimizing the access policy, which selects the number of repetitions, is a complicated problem, and failing to do so can lead to an inefficient use of resources and, potentially, to an increased congestion. In this paper, we follow a game-theoretic approach for optimizing the access policies of selfish users in modern random access mechanisms. Our goal is to find adequate values for the rewards given after a success to achieve a \gls{ne} that optimizes the throughput of the system while considering the cost of transmission. 
Our results show that a \emph{mixed strategy}, where repetitions are selected according to the \gls{irsa} protocol, attains a \gls{ne} that maximizes the throughput in the special case with two  users. In this scenario, our method increases the throughput by $30$\% when compared to framed ALOHA. Furthermore, we present three methods to attain a \gls{ne} with near-optimal throughput for general modern random access scenarios, which exceed the throughput of framed ALOHA by up to $34$\%.



\end{abstract}
\glsresetall
\glsresetall

\section{Introduction}
The number of mobile devices is increasing rapidly, which has raised concerns about the capacity of the access channels of cellular networks. In particular, the random access mechanisms of 4G and 5G, including \gls{nbiot}, 
are based on a simple implementation of multichannel slotted ALOHA. Therefore, the access channels of these systems have a capacity (i.e., average throughput per slot) of $R/e$ packets per time slot, with $R$ being the number of available channels. This capacity is insufficient to support the massive number of \gls{iot} users that are expected to be deployed in the following years~\cite{Nguyen22}. Hence, one of the main goals of 6G is to design novel random access mechanisms with a much higher capacity and resource efficiency than those in 5G.

\begin{figure}[t]
\centering
\includegraphics{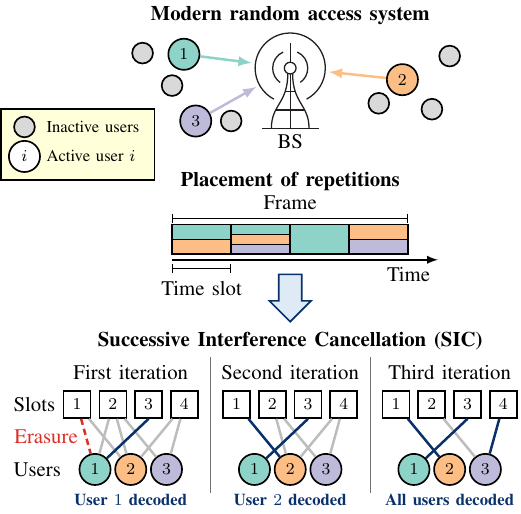}
\vspace{-20pt}
\caption{Exemplary modern random access system where $N=3$ active users transmit repetitions of the same packet according to \gls{irsa} and the BS employs SIC for multi-user detection.}
\label{fig:example}
\vspace{-10pt}
\end{figure}

Modern random access mechanisms are a promising solution to improve the capacity and the resource efficiency of wireless channels by combining packet repetitions and advanced multi-user decoding techniques. A prominent example of modern random access is \gls{irsa}~\cite{Liva11}, where the users select the number of repetitions to transmit within a time interval randomly according to a degree distribution. Then, these repetitions are placed uniformly at random within the interval. The randomness in the number and placement of the repetitions in \gls{irsa}, in combination with \gls{sic},  provides an advantageous temporal diversity to the scheme that can significantly reduce the error probability and increase the throughput of the system when compared to traditional slotted ALOHA. Fig.~\ref{fig:example} illustrates the operation of \gls{irsa} with $N=3$ active users, including the placement of repetitions and \gls{sic}, where the outcome is the decoding of all the three transmitted packets.  

Despite its benefits, the randomness of \gls{irsa} limits its analytical tractability and complicates its optimization. A viable approach to optimize the access policy of the user is to model the problem as a non-cooperative game, where each user acts in a selfish manner for obtaining its optimal policy. For this, the users must consider the possible actions of the other players to find the policy that maximizes their own \emph{utility}, which is typically calculated both in game theory and reinforcement learning literature, as the reward for delivering the user data minus the cost of transmissions. If there exists a point where no user can achieve a bigger utility by changing their own policy, a \gls{ne} is reached and the policy of the users is stable.

A further complication of modern random access is the reliance of \gls{sic} on performing accurate channel estimation, even for repetitions that were received with a \gls{sinr} that is below the threshold for decoding. Therefore, analyses that assume a perfect decoding of singleton slots only provide an upper bound in performance. To better characterize the performance of \gls{irsa}, the upper bound under the assumption of perfect decoding can be accompanied with a lower bound, for example, by assuming a case where a singleton slot can not be decoded.

In this paper, we present a game-theoretic analysis of \gls{irsa}, where we model the random access process as a non-cooperative game with incentives. Our analyses focus on determining the numerical values for the incentives that lead to Nash equilibria that maximize the throughput of the system. This allows the users to derive access policies that both optimize the access of the individual user and the performance of the \gls{irsa} system in a distributed and selfish manner. 
We consider multiple scenarios ranging from a two-user contention case, scenarios with short frames, as a well as a massive access scenario. 
The main contribution of this paper is threefold. First, we demonstrate that the mixed strategies used in \gls{irsa} lead to a \gls{ne} that maximizes throughput in the two-user case. Second, we demonstrate a framework to determine whether the optimal degree distribution obtained in the asymptotic case, where the number of users and frame length tend to infinity, can be a \gls{ne} in practical systems. Third, we present two methods to define the rewards that lead to a \gls{ne} with near-optimal throughput. 
 

\section{Related work} 
Optimizing access policies in modern random access is a complicated endeavor. Hence, numerous studies focus on the performance analysis and optimization of modern random access schemes in the asymptotic case~\cite{Ceda18, Liva11, diffevo}. Among these, \emph{differential evolution} can be used to obtain a degree distribution that maximizes the throughput of the system~\cite{diffevo}. However, the performance of degree distributions obtained in the asymptotic case degrades significantly in systems with short frames and user populations. Therefore, other studies rely on the use of graph theory to derive approximations for the packet loss rate with reduced frame lengths~\cite{Moroglu20}. However, these focus on the performance evaluation and do not provide mechanisms to optimize the degree distributions.


Conversely,  only a few studies have applied game theory on \gls{irsa} settings with selfish users (i.e., players). It has been observed that selfish users tend to transmit the maximum number of repetitions allowed to maximize their reliability~\cite{Clazzer18}. Nevertheless, the existence of \gls{ne} in ALOHA-based systems has been proven~\cite{Hmedoush22}. 
Furthermore, it has been observed that an efficient \gls{ne} can be attained in framed ALOHA systems that focus on reducing \gls{aoi}~\cite{Badia24}.
In addition, an achievable payoff region is obtained, along with its \gls{ne}, in a two-user non-cooperative slotted ALOHA game~\cite{seo18}.
However, the numerical values for the reward given to the users, which is the main focus of this paper, have not been studied previously. 


  \section{System model}\label{sec:sys_mod}
We consider a random access problem in a wireless system where users communicate sporadically over a time-slotted channel towards a \gls{bs}. The operation of the system is divided into frames of $M$ time slots and each packet transmission occupies a whole time slot. We focus on the analysis of individual frames with $N$ active users, where the system load is $G=N/M$ users per slot.

The users communicate with the base station following an \gls{irsa} protocol, where the number of repetitions of the same packet transmitted within a frame $d$ is selected randomly according to a \emph{degree distribution}. The polynomial representation of the degree distribution is
\begin{equation}
    \Lambda(x):=\sum_{d=1}^{d_{max}}\Lambda_d x^d,
\end{equation}
where $\Lambda_d$ is the probability of transmitting $d$ repetitions in a frame. Note that \gls{irsa} is a generalization of modern random access mechanisms based on repetitions, as mechanisms considering a fixed number of repetitions are a special case of \gls{irsa} where $\exists \Lambda_d=1$. 

Communication takes place over a collision channel with erasures, where the probability of decoding a packet transmitted by a user in a slot with interference from other users is $0$. Furthermore, the probability of decoding a packet that is not affected by interference (i.e., in a singleton slot) is $p$, which is set to an equal value for all the users. Thus, $1-p$ is the erasure probability. Note that the collision channel is a pessimistic scenario for \gls{irsa} and \gls{sic}, as the performance of the scheme would increase in a channel with capture~\cite{Clazzer17}. 

After receiving the signals, the \gls{bs} attempts to decode the packets without collision. The repetitions include pointers that enable the \gls{bs} to identify all the slots with repetitions of the same packet upon the successful decoding of one or more of them. These pointers are used to perform \gls{sic} across the slots, which is assumed to perfectly remove the interference from the repetitions of the same packet.

As shown at the bottom of Fig.~\ref{fig:example}, the operation of \gls{irsa} can be modeled as a bipartite graph, with the vertex set including the user- and the time slot-nodes~\cite{Liva11}. Some of these nodes are connected by edges, indicating the transmission of a repetition by a given user in a given time slot. We define the user- and the slot-node degree distributions, respectively, as
\begin{equation*}
    \Lambda(x) := \sum_{d=1}^{d_{max}} \Lambda_dx^d, \quad P(x) := \sum_{c=1}^{N} P_cx^c.
\end{equation*}
Next, let $\lambda_h$ be the probability that an edge is connected to a user-node with degree $h$ and $\rho_h$ be the probability that an edge is connected to a slot-node with degree $h$. These are given by the polynomial representations 
\begin{equation*}
    \lambda(x) := \!\sum_{d=1}^{d_{max}}\! \lambda_dx^{d-1} = \frac{\Lambda'(x)}{\Lambda'(1)}, \rho(x) := \!\sum_{c=1}^{N} \!\rho_cx^{c-1}=\frac{P'(x)}{P'(1)},
\end{equation*}
related to the user- and the slot-node degree distributions as
\begin{equation*}
    \lambda_h = \frac{h\Lambda_h}{\displaystyle \sum_{d=1}^{d_{max}}d\Lambda_d}, \quad
    \rho_h = \frac{hP_h}{\displaystyle \sum_{d=1}^{d_{max}}dP_d}.
\end{equation*}

Let $q_i$ be the probability that an edge is unresolved after $i$ \gls{sic} iterations. Next, let $\varrho_{i,h}$ be the probability that the edge emanating from a degree $h$ slot node is unresolved after $i$ \gls{sic} iterations. Then, for a user transmitting $d$ repetitions we have $q_i = \varrho_{i,h}^{d-1}$. 
Taking the average over all possible edge-oriented degrees, the probability of an edge being unresolved in the $i$-th \gls{sic} iteration is determined by the recursive equations: 
\begin{align}
    q_i &:= 
    \sum_{d=1}^{d_{max}} \lambda_d \varrho_{i-1,h}^{d-1} = \lambda\left(\varrho_{i-1,h}\right)\\
    \varrho_{i,h} &:= 
    \sum_{h=1}^{N}\rho_h \left(1-(1-q_{i})^{h-1}\right) = 1-\rho\left(1- q_{i}\right),
\end{align}
 Since all packets are unresolved before the first iteration, the initial condition is $q_0 = 1$. Thus, $\varrho_{i,h} = 1- \text{e}^{-q_iG\Lambda'(1)}$.


By definition, $\varrho_{\infty,h}^d$ is the probability that all the $d$ packets of a user are unresolved after infinitely many SIC iterations. 
In such case, the ratio between unresolved users and the total number of users is denoted as the \gls{plr}, which is defined as a function of a degree distribution as 
\begin{equation}
P_L(\Lambda):=\sum_{d=1}^{d_{max}}\Lambda_d\varrho_{\infty,h}^{d} = \Lambda\left(\varrho_{\infty,h}\right).
\end{equation}

This leads to the definition of throughput per slot, which is the main performance indicator used throughout this paper, as
\begin{equation}
    T(G,\Lambda):= G(1-P_L(\Lambda)).
\end{equation}

\section{A game-theoretic approach to user rewards}
\label{sec:approach}
In this section, we present game-theoretic approach to maximize the performance of a modern random access system with selfish users. As in other game-theoretic and learning frameworks, the users (i.e., players) will take actions following a \emph{strategy profile} and receive a reward for successful access while incurring in a cost $c$ for their access attempts. In our case, the \emph{strategy profile} is the policy followed by the players to choose the number of repetitions. Our goal is to fine-tune the rewards given to the users in a general modern random access system to attain a \gls{ne} that achieves a near-optimal throughput. This is the ideal scenario for the system with selfish users, since they will have no motivation to use a strategy that is different to the one that benefits the performance of the whole system. 


In the remainder of this section, we present four methods to attain a \gls{ne} with near-optimal throughput. These are based on the following theorem from game theory describing the conditions for the \gls{ne}. 

\newtheorem{theorem}{Theorem}

\begin{theorem}[Conditions for the Nash Equilibrium]\label{n_nash}\textbf{}\\
    A mixed strategy profile $\mathbf{s}^*\! \in\! S$ is a NE if, and only if, the following two conditions hold for any player $i$~\cite[p.71]{game_wireless}.
    
   \noindent1) The utility of all pure strategies in the support $\delta(s_i^*)$ will be equal. Namely,\begin{equation}
            u_i\left([a_i,\mathbf{s}^*_{-i}]\right)=u_i\left([b_i,\mathbf{s}^*_{-i}]\right), \forall a_i,b_i\in \delta(s^*_i).
        \end{equation}
   2) The utility of any pure strategy in the support $\delta(s_i^*)$ is greater than or equal to that of any $b_i\notin\delta(s_i^*)$. Namely,
        \begin{equation}
        u_i\left([a_i,\mathbf{s}^*_{-i}]\right)\geq u_i\left([b_i,\mathbf{s}^*_{-i}]\right) \hspace{0.1em}  \forall a_i\in \delta(s^*_i)  \wedge \forall b_i\notin \delta(s^*_i).
        \end{equation}
The proof is found in the Appendix.

\end{theorem}

\subsection{\Gls{ne} with two users and no erasures}
\label{sec:two_users}
In this simplified scenario, we are interested on determining whether the \emph{strategy profile} that leads to the \gls{ne} with maximum throughput is a \emph{mixed strategy}. This is of especial relevance, as a mixed strategy corresponds to the \gls{irsa} scheme, whereas a constant number of repetitions  corresponds to a pure strategy.

The throughput of a system depends on the load and ratio of resolved users. Thus, for a system with a constant load $G$, the throughput is only dependent on the ratio of successful users. Thus, maximizing the \gls{esu} is the same as maximizing the throughput. 
Let $p_{s_n}([i,j])$  be the probability of a successful transmission by user $n\in\{1,2\}$
for a strategy profile $[i,j]$, where $i$ and $j$ are the number of repetitions performed by the each user. Then, we define the \gls{esu} as
\begin{equation}
E([i,j]) = p_{s_1}([i,j]) + p_{s_2}([i,j])\in[0,2]. 
\label{eq:esu_2}
\end{equation}

Since we assume a perfect SIC procedure, the only case where both users do not successfully transmit is when all transmitted packets collide. 
Therefore, their probability for a successful transmission is given by 
\begin{equation}\label{eq:SIC_success}
    p_{s_n}\left([i,j]\right ) = \begin{cases}
        1, \quad &\text{for } i \neq j, \\    
    1-\frac{1}{\binom{M}{i}}, \quad &\text{for } i = j,  
    \end{cases}\quad \text{for } n \in \{1,2\},
\end{equation}
which is maximized when the users play different strategies, namely, $i\neq j$. Then, the \gls{esu} is $E([i,j])=2p_{s_n}\left([i,j]\right )$.

To find the mixed strategy \gls{ne}, we apply Theorem \ref{n_nash} to define a system of $M$ equations with $M$ variables. From condition 1 in Theorem \ref{n_nash}, we want a mixed strategy $s_1 = s_2$ where $E([s_1,s_2]) = E([i,s_2])$ for each $i \in \{1,...,M\}$.  
Due to condition 1, the solution to the system of equations will be a mixed strategy \gls{ne} that maximizes the throughput, s.t. all pure strategies in the strategy space are in its support. 
Thus, the strategy profile that maximizes the throughput of the system assigns the distribution $\{\Lambda_1,\dotsc,\Lambda_M\}$ to both players, where 
\begin{equation}\label{eq:distribtuin_k}
    \Lambda_k = \frac{M-(k-1)}{k}\Lambda_{k-1} =  \prod_{i=1}^{k-1}\left(M-(k-i)\right)\frac{\Lambda_1}{k!},
\end{equation} for $k \in \{2,...,M\}$, and 
\begin{equation}\label{eq:distribtuin_1}
\Lambda_1 = 1-\sum_{k=2}^M \Lambda_k = \left(1+\sum_{k=2}^M \prod_{i=1}^{k-1} \frac{M-(k-i)}{k!}\right)^{-1}\hspace{-5pt}.
\end{equation}

By letting $r\to\infty$ and considering the utility function
\begin{equation}
    u_1\left([i,j]\right) = p_{s_1}\left([i,j]\right)r - ic= \frac{1}{2}E\left([i,j]\right)r - ic,
\end{equation} 
the cost $c$ will have a negligible effect on the utility of the system. Furthermore, scaling the \gls{esu} function does not change the strategy profile that maximizes it. Therefore, by choosing a sufficiently high reward $r\to\infty$, the \gls{ne} tends towards the mixed strategy profile that maximizes the throughput of the system, given by \eqref{eq:distribtuin_k} and \eqref{eq:distribtuin_1}.

\subsection{General case with $N$ users}

In the following, we present a game-theoretic analysis towards obtaining the strategies that lead to a \gls{ne} that maximizes the throughput of the system for general case with $N$ users and $M$ time slots. 
First, we present a method to demonstrate the feasibility of achieving a \gls{ne} with a degree distribution obtained in using differential evolution \cite{diffevo} the asymptotic case with $N, M\to\infty$. Then, we present a method that employs a constant reward to derive the degree distribution that maximizes throughput in scenarios with short frame lengths. Finally, we present an iterative method to achieve a \gls{ne} with a near-optimal degree distribution.


\subsubsection{\Gls{ne} with differential evolution} 
Following the conditions in Theorem \ref{n_nash}, we aim to identify the utility function that imposes a \gls{ne} with the degree distribution obtained by differential evolution in the asymptotic case where $M, N\to\infty$~\cite{diffevo}, denoted as $\Lambda_\infty(x)=\sum_{d=1}^{d_\text{max}} \Lambda_d^\infty x^d$. Thus, we define

\begin{IEEEeqnarray}{rCl}
u_i\left(\left[d,\mathbf{s}_{-i}\right]\right)&=&r_d \,p_{s_i}\left(\left[d,\mathbf{s}_{-i}\right]\right) - d\,c, \forall d\in\{1,\dotsc,d_\text{max}\}\IEEEeqnarraynumspace\\
    u_i\left(\left[{s}_i,\mathbf{s}_{-i}\right]\right) &=&r_\infty \, p_{s_i}\left(\left[{s}_i,\mathbf{s}_{-i}\right]\right) - \sum_{d=1}^{d_{max}} dc \Lambda_d^\infty,
\end{IEEEeqnarray}
where $r_{\infty}=\sum_{d=1}^{d_{max}}\Lambda_d^\infty \cdot r_d$. The values for $p_s$ must be found by Monte Carlo simulation with the precise values of $N$ and $M$ for the scenario, where $N-1$ users follow the  degree distribution $\Lambda_\infty(x)$, but one user deviates by following a pure strategy.
The system of equations that establishes our framework for making the optimal strategy a \gls{ne} is
\begin{IEEEeqnarray}{rCl}
u_i\left(\left[d,\mathbf{s}_{-i}\right]\right)
&=&u_i\left(\left[{s}_i,\mathbf{s}_{-i}\right]\right), \quad \forall d\in\delta(s_i)\\
u_i\left(\left[d,\mathbf{s}_{-i}\right]\right)
&\leq &u_i\left(\left[{s}_i,\mathbf{s}_{-i}\right]\right), \quad \forall d\notin\delta(s_i)
\label{eq:se_diff_evo}
\end{IEEEeqnarray}

This method assumes that the reward can vary depending on the number of repetitions that a user transmits. However, most learning mechanisms employ a constant reward for each successful access. Therefore, in the ideal case, the previous method would result in a constant reward $r_1=r_2,\dotsc,r_\text{max}$, which can be implemented to achieve a \gls{ne} in practical scenarios. Otherwise, the reward mechanism would require adaptations to make the distribution $\Lambda_\infty(x)$ a \gls{ne}. 
\subsubsection{Short-frame \gls{irsa} \gls{ne}}
In the following, we describe a process to attain a \gls{ne} with a constant reward $r$, which follows the conventions of most learning frameworks. However, due to its computational complexity, this method is restricted to systems with short frames.

We first define the expression for the utility of user $i$ when transmitting $d$ packets by considering all the combinations of what the $N-1$ other users can transmit, as
\begin{multline*}
u_i([d,\mathbf{s}_{-i}])= \\ 
\sum_{\substack{k_1,...,k_{d_{max}}\geq0 \\ k_1+\cdots+k_{d_{max}} = N-1}}^{N-1} \frac{(N-1)!}{k_1!\cdots k_{d_{max}}!} \prod_{t=1}^{{d_{max}}}\Lambda_t^{k_t}u_i\left([d, h(k_t,t)]\right),
\end{multline*}
where $h(k_t,t)$ indicates there are $k_t$ users transmitting $t$ packets. Next, we select the support for the degree distribution $d\in\delta(s_i^*)$ and define a system of equations to impose a \gls{ne} with a mixed strategy that includes all the pure strategies in the support $\delta(s_i^*)$, as 
\begin{IEEEeqnarray}{rl}
    u_i\left(\left[d,\mathbf{s}_{-i}\right]\right) \,&=u_i\left(\left[d',\mathbf{s}_{-i}\right]\right),\quad \forall d,d'\in\delta(s_i^*)\label{eq:trans_two}\\
    u_i\left(\left[d,\mathbf{s}_{-i}\right]\right) \,&\geq u_i\left(\left[d',\mathbf{s}_{-i}\right]\right),\quad \forall d\in\delta(s_i^*), d'\notin\delta(s_i^*)\IEEEeqnarraynumspace\\
\sum_{d=1}^{d_\text{max}}\Lambda_d\, &= 1.
\end{IEEEeqnarray}
Since this method yields a system of $d_{max}$ nonlinear equations it is difficult to determine a solution when either $N$ or $d_\text{max}\leq M$ increase. A further complication of this method is that it requires the \emph{a-priori} selection of the support $\delta(s_i^*)$, which might result in sub-optimal degree distributions.

\subsubsection{Best reply}
By definition, no user in a \gls{ne} can change their strategy to get a higher utility. Thus, if all users keep improving their utility until they can no longer do so, a \gls{ne} has been reached. This definition is used to obtain the degree distribution that yields a \gls{ne} by simulation for any $N$, $M$, $r$ and $p$, the probability of decoding a packet without collision. 

This method, called ``Best reply,'' is initialized by defining the values for $N$, $M$, $d_\text{max}$, and $r$ while setting all users to transmit $1$ repetition, which corresponds to the framed slotted ALOHA access protocol, that can then be improved. 
The goal is to determine the
\gls{ne} by updating the utility of a specific user $i$ while
fixing the number of transmissions from every other user. Thus, user $i$ transmits the
number of packets that yields the highest average utility over a large number of simulated frames and its degree distribution is updated. Then, a different user is considered and the same process is repeated until every user has optimized their utility once. Then, the initial user $i$ selected and this process is done iteratively
until the utility of the users and their degree distributions remain constant between two iterations. Thus, a \gls{ne} has been determined.


\section{Results}




This section presents numerical results obtained with the game-theoretic methods described previously. In all cases, we set the transmission cost to $c=1$ and consider framed ALOHA (i.e., $\Lambda(x)=x$) for benchmark purposes.

First, we exemplify the benefits of our game-theoretic approach described in Section~\ref{sec:two_users} by identifying a \gls{ne} with two users and no erasures for $M=4$. In this case, by setting a sufficiently high reward, namely $r=40000$, the optimal degree distribution converges to $\Lambda(x)=0.27x+0.4x^2+0.26x^3+0.07x^4$. The latter yields a throughput of $0.49$, which is $30$\% higher than the one achieved with framed ALOHA.

We proceed to general cases with $N$ users by presenting  the degree distributions obtained in the asymptotic case with differential evolution~\cite{diffevo} for $G=0.9$ and $d_{max}\in\{2,3,\dots,6\}$ in Table~\ref{tab:diff_evo}. Clearly, the performance of the scheme increases with $d_{max}$ in the asymptotic case. However, the $d_{max}$ is restricted by the frame length $M\geq d_{max}$ in practical scenarios and, as it will be seen in the following results, the performance of the degree distributions obtained with differential evolution greatly decreases for $M\leq 100$. 

\begin{table}[t]
\centering
\caption{Degree distributions and  throughput per slot $T(G,\Lambda)$ obtained in the asymptotic case using differential evolution~\cite{diffevo}.}\vspace{-6pt}
\renewcommand{\arraystretch}{1.2}
\begin{tabular}{|c|c|c|c|}
 \hline
 $d_{max}$ & Degree Distribution $\Lambda_\infty(x)$ & Avg. Repetitions & 
 $T(G,\Lambda)$ \\ 
 \hline
 $2$ & $x^2$ & $2$ & $0.543$    \\ 
 \hline
 $3$& $0.10x  + 0.90x^3$ & $2.8$ & $0.835$    \\
 \hline
 $4$&$0.51x^2 +0.49x^4$ &  $2.98$ &  $0.864$    \\
 \hline
 $5$& $0.55x^2 + 0.06x^3+0.39x^5$ &  $3.23$ &  $0.878$    \\
 \hline
 $6$& $0.54x^2 + 0.17x^3 + 0.29x^6$ &  $3.33$ &  $0.910$\\
 \hline
\end{tabular}
\label{tab:diff_evo}
\vspace{-0.3cm}
\end{table}

Next, we aim to obtain the degree distributions that lead to a \gls{ne} in a short-frame scenario with $M=5$ and $G = 0.8$. As described in Section~\ref{sec:approach}, our method for short frame \gls{irsa} \gls{ne} takes the support for the degree distribution $\delta(s_i^*)$ as input to find  the degree distribution $\Lambda(x)$, which has a major impact on the achieved throughput. Table~\ref{tab:short_frame} shows the degree distributions and the per-slot throughput obtained by selecting different supports and setting the reward to a relatively high value $r=500$. Among the selected sets for $\delta(s_i^*)$, the highest throughput is achieved with $\delta(s_i^*)=\{1,2,3\}$, which is $34\%$ higher than with framed ALOHA and $71\%$ higher than with differential evolution for $d_{max}=4$. While this illustrates the potential benefits of our approach, Table~\ref{tab:short_frame} also shows that it is greatly sensitive to the support $\delta(s_i^*)$, as  the throughput drops significantly to $0.3$ by simply including $d=4$ in $\delta(s_i^*)$.

\begin{table}[t]
\centering
\caption{Degree distributions and  throughput per slot $T(G,\Lambda)$ obtained in the short-frame scenario with $M=5$ and $G=0.8$.}\vspace{-6pt}
\renewcommand{\arraystretch}{1.2}
\begin{tabular}{|c|c|c|}
 \hline
 Support $\delta(s_i^*)$ & Degree Distribution $\Lambda(x)$ &$T(G,\Lambda)$  \\ 
 \hline
 Framed ALOHA & $x$ & $0.41$    \\ 
 \hline
 $\{2,3\}$ & $0.54 x^2 + 0.46 x^3 $ & $0.44$    \\
 \hline
  $\{1,2,3\}$ &$0.25 x + 0.38 x^2 + 0.37 x^3$  &  $0.55$    \\
 \hline
  $\{1,2,3,4\}$ & $0.03x+0.249x^2+0.4x^3+0.321x^4$ &   $0.30$\\
 \hline
 $\Lambda_\infty(x)$ with & $0.51x^2+0.49x^4$ &$0.32$\\[-3pt] 
$d_{max}=4$ & & \\
 \hline
\end{tabular}
\label{tab:short_frame}
\vspace{-0.3cm}
\end{table}

Finally, we consider the scenario with $M=100$ and $G=0.9$. 
First, we aim to determine whether a constant reward can be used to impose a \gls{ne} in the system using the degree distribution $\Lambda_\infty(x)$ obtained by setting $d_{max}=6$ with differential evolution.
We begin by allowing a varying reward depending on the number of packets transmitted and estimating
the success probabilities by Monte Carlo simulation. These are used to obtain the system of equations in~\eqref{eq:se_diff_evo}, which results in the following values for the reward that induce $\Lambda_\infty(x)$ to be the desired \gls{ne}: $r_1\leq19.79$, $r_2=17.26$, \mbox{$r_3=16.58$,}
 $r_4\leq16.63$, $r_5\leq16.86$, 
    $r_6=17.39$.

The inequalities for $r_1$, $r_4$, and $r_5$ are introduced since $d\in\{1, 4,5\}$    repetitions is not in the support of  $\Lambda_\infty(x)$. As it can be seen, the rewards for the three pure strategies in the support of $\Lambda_\infty(x)$ must be different to induce the desired \gls{ne}. Hence, the latter cannot be achieved by using a constant reward $r$.

Therefore, to achieve a \gls{ne} with a single constant reward, we apply the Best reply method described in the previous section for $d_{max}=6$. At each iteration, $10000$ frames were simulated for each user. By setting the reward  $r=8$, we observed that an equal utility of $1.1$ is obtained for $d\in\{2,3\}$ and the utility is lower for all other values of $d$.
Therefore, the degree distribution found by Best reply to yield a \gls{ne} is 
    $\Lambda(x) = 0.5x^2 + 0.5x^3$,
which results in a \gls{plr} of $0.55$ and a throughput of $0.405$. While this configuration exceeds the throughput of framed ALOHA by $10$\%, finding the \gls{ne} that maximizes  throughput with Best reply requires searching through the possible values of $r$.

\begin{figure}[!t]
\centering
\subfloat[]{\includegraphics{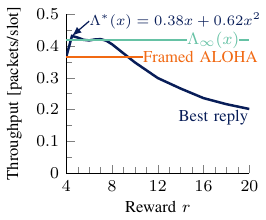}}\hfil\subfloat[]{\includegraphics{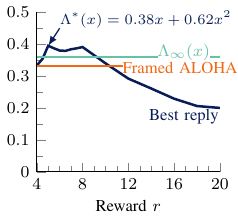}}
\caption{Throughput using Best reply given $M=100$ and $G=0.9$ for different rewards $r$ and decoding probabilities (a) $p=1$ and (b) $p=0.9$. The throughput with framed ALOHA and the distribution obtained with differential evolution in the asymptotic case $\Lambda_\infty(x)$ are included as benchmarks.}
\vspace{-0.2cm}
\label{fig:sim_results}
\end{figure}

Thus, Fig.~\ref{fig:sim_results} shows the throughput of the degree distributions that yield a \gls{ne} for  $r\in\left[4,20\right]$ and $p=\{0.9,1\}$. In addition, the throughput of framed ALOHA and with the optimal asymptotic distribution $\Lambda_\infty(x) = 0.54x^2 + 0.17x^3 + 0.29x^6$, which are insensitive to the reward $r$, are included as a reference. As it can be seen, the highest throughput is achieved with Best reply for both values of $p$. Namely, the throughput with Best reply and $p=1$ is up to
$17$\% higher than with framed ALOHA and up to $3$\% higher than with $\Lambda_\infty(x)$. Furthermore, the throughput with Best reply and $p=0.9$ is up to
$11$\% higher than with framed ALOHA and up to $10$\% higher than with $\Lambda_\infty(x)$. The reason for the increased difference between Best reply and $\Lambda_\infty(x)$ for $p=0.9$ is that distribution with the former was obtained by directly simulating the system with the exact value of $p$, whereas the distribution obtained with differential evolution $\Lambda_\infty(x)$ assumes $p=1$.


Furthermore, we see that the highest throughput for $p=1$ is achieved by setting $r=4.5$ and for $p=0.9$ by setting $r=5$. The reason for this is that the number of number of repetitions yielding a \gls{ne} increases along with the reward. For instance, by setting a reward $r\leq 4$, the resulting degree distribution is $\Lambda(x)=x$, the same as framed ALOHA.  It is also seen that the throughput decreases when $r\geq 9$. This behavior is a case of the \emph{tragedy of the commons}, where the high reward given for each successful access results in greedy players (i.e., users) transmitting too many repetitions in the frames. Therefore, it exemplifies the need for selecting an adequate reward $r$.


\section{Conclusions}
In this paper, we adopted a game-theoretic approach to modern random access. We provided four methods to derive the values of the incentives to be given to the users so the \gls{ne} leads to a maximum system throughput, which are applicable to multiple scenarios, defined by the number of active users and the decoding probability. 
For a system restricted to two users, it is proven that a mixed strategy optimizes the throughput. Then, in a system with short frames, it is shown that the throughput can be improved by $34$\% when compared to framed ALOHA and by $71$\% when compared to differential evolution. However, minor variations in the support for the degree distribution, which must be selected \emph{a priori}, can have a profound negative impact on throughput. Finally,
in a  system with longer frames, our Best reply method outperforms framed ALOHA and the distribution obtained by differential evolution by up to $17\%$. However, Best reply method is greatly sensitive to the value of the reward, which must be selected beforehand. Thus, our future work will focus on learning mechanisms for the configuration of our methods to maximize  throughput.


\section{Appendix}\label{sec:prof}

\begin{proof}
This proof for Theorem~\ref{n_nash} is inspired by \cite[pp. 8-9]{prof_thm}.
Let $\mathbf{s}^*$ be a NE. 
By definition, we have
\begin{equation*}
u_i\left([s^*_i,\mathbf{s}^*_{-i}]\right)\geq u_i\left([s_i,\mathbf{s}^*_{-i}]\right), \quad\forall s_i \in S_i,
\end{equation*}
so that
\begin{equation*}
    u_i\left([s^*_i,\mathbf{s}^*_{-i}]\right) = \max_{s_i\in S_i}u_i\left([s_i,\mathbf{s}^*_{-i}]\right).
\end{equation*}
The expected utility for a mixed strategy profile can be written as a convex combination using the utilities from the pure strategies. Thus, it follows that
\begin{align*}   
u_i\left([s^*_i,\mathbf{s}^*_{-i}]\right) 
    &= \max_{s_i\in S_i}\sum_{a_i \in A_i} s_i(a_i) u_i \left([a_i,\mathbf{s}^*_{-i}]\right)  \\
    &= \max_{a_i\in A_i}u_i\left([a_i,\mathbf{s}^*_{-i}]\right).
\end{align*}
The utility $u_i([s_i^*,\mathbf{s}_{-i}^*])$ can be written as a convex combination of the pure strategies in the support, since $s^*_i(a_i)=0$ $\forall a_i \notin \delta(s^*_i)$, resulting in
\begin{equation*}
   \sum_{a_i \in \delta(s^*_i)} s^*_i(a_i)u_i\left([a_i,\mathbf{s}^*_{-i}]\right) = \max_{a_i\in A_i}u_i\left([a_i,\mathbf{s}^*_{-i}]\right),
\end{equation*}
which implies that 
  $u_i\left([a_i, \mathbf{s}^*_{-i}]\right)=u_i\left([b_i, \mathbf{s}^*_{-i}]\right), \quad \forall a_i, b_i \in \delta(s^*_i)$.
This leads to \[u_i([a_i, \mathbf{s}^*_{-i}])=\max_{\hat{a}_i\in A_i}u_i\left([\hat{a}_i,\mathbf{s}^*_{-i}]\right), 
\forall a_i \in \delta(s^*_i).\]
Therefore
\begin{equation*}
    u_i\left([s^*_i,\mathbf{s}^*_{-i}]\right)=u_i\left([a_i,\mathbf{s}^*_{-i}]\right),\quad \forall a_i \in \delta(s^*_i)
\end{equation*}
and 
\begin{equation*}
    u_i\left([a_i,\mathbf{s}^*_{-i}]\right)\geq u_i\left([b_i,\mathbf{s}^*_{-i}]\right),\quad \forall b_i \notin \delta(s^*_i) ~\wedge~ \forall a_i \in \delta(s^*_i).
\end{equation*}

\hspace*{\fill}\LARGE$ \diamond$ \normalsize\\
We assume now that a mixed strategy profile $\mathbf{s}^*$ satisfies the conditions 1 and 2.
First, we use condition 1 to define $w_i:=u_i([a_i,\mathbf{s}^*_i]), \forall a_i \in \delta(s^*_i).$ Then, by writing $u_i([s^*_i,\mathbf{s}^*_{-i}])$ as a convex combination, it follows that
\begin{align}
    u_i\left([s^*_i,\mathbf{s}^*_{-i}]\right)&=\sum_{a_i\in \delta(s^*_i)}s^*_i(a_i)u_i\left([a_i,\mathbf{s}^*_{-i}]\right)\nonumber\\&=\sum_{a_i\in \delta(s^*_i)}s^*_i(a_i)w_i=w_i.\label{eq:weq}
\end{align}
Similarly, with condition 2 we get 
\begin{align}
    w_i=\sum_{a_i\in A_i} s_i(a_i)w_i&\geq \sum_{a_i\in A_i} s_i(a_i)u_i\left([a_i,\mathbf{s}^*_{-i}]\right)\nonumber\\ &=u_i\left([s_i,\mathbf{s}^*_{-i}]\right),  \quad \forall s_i \in S_i.\label{eq:wgeq}
\end{align}
Combining Equation (\ref{eq:weq}) and Equation (\ref{eq:wgeq}) results in
\begin{equation*}
    u_i\left([s^*_i,\mathbf{s}^*_{-i}]\right) \geq u_i\left([s_i,\mathbf{s}^*_{-i}]\right), \quad \forall s_i \in S_i.
\end{equation*}
\end{proof}

\bibliographystyle{IEEEtran}
\bibliography{bib}
\end{document}